\documentclass[aps,prb,reprint,superscriptaddress,floatfix]{revtex4-2}
\usepackage[normalem]{ulem}
\usepackage{xcolor}
\usepackage{graphicx}
\usepackage{amsmath}
\usepackage{physics}
\usepackage[colorlinks=true,allcolors=blue]{hyperref}%

\begin{document}


\title{Tendencies of enhanced electronic nematicity in the Hubbard model and a comparison with Raman scattering on high-temperature superconductors}

\author{Tianyi Liu}
\affiliation{Stanford Institute for Materials and Energy Sciences,
SLAC National Accelerator Laboratory, 2575 Sand Hill Road, Menlo Park, CA 94025, USA
}
\affiliation{Department of Chemistry, Stanford University, Stanford, CA
94305, USA}
\author{Daniel Jost}
\affiliation{Walther Meissner Institut, Bayerische Akademie der Wissenschaften, 85748 Garching, Germany
}
\author{Brian Moritz}
\affiliation{Stanford Institute for Materials and Energy Sciences,
SLAC National Accelerator Laboratory, 2575 Sand Hill Road, Menlo Park, CA 94025, USA
}
\affiliation{Department of Physics and Astrophysics, University of North Dakota, Grand Forks, North Dakota, 58202, USA}
\author{Edwin W. Huang}
\affiliation{Department of Physics and Institute of Condensed Matter Theory,
University of Illinois at Urbana-Champaign, Urbana, IL 61801, USA.
}
\author{Rudi Hackl}
\affiliation{Walther Meissner Institut, Bayerische Akademie der Wissenschaften, 85748 Garching, Germany
}
\author{Thomas P. Devereaux}
\affiliation{Stanford Institute for Materials and Energy Sciences,
SLAC National Accelerator Laboratory, 2575 Sand Hill Road, Menlo Park, CA 94025, USA
}
\affiliation{Department of Materials Science and Engineering, Stanford University, Stanford, CA 94305, USA}


\date{\today}

\begin{abstract}
The pseudogap regime of the cuprate high-temperature superconductors is characterized by a variety of competing orders, the nature of which are still widely debated. Recent experiments have provided evidence for electron nematic order, in which the electron fluid breaks rotational symmetry while preserving translational invariance. Raman spectroscopy, with its ability to symmetry resolve low energy excitations, is a unique tool that can be used to assess nematic fluctuations and nematic ordering tendencies.  Here, we compare results from determinant quantum Monte Carlo simulations of the Hubbard model to experimental results from Raman spectroscopy in La\textsubscript{2-x}Sr\textsubscript{x}CuO\textsubscript{4}, which show a prominent increase in the $B_{1g}$ response around 10\% hole doping as the temperature decreases, indicative of a rise in nematic fluctuations at low energy. Our results support a picture of nematic fluctuations with $B_{1g}$ symmetry occurring in underdoped cuprates, which may arise from melted stripes at elevated temperatures.
\end{abstract}


\maketitle

\section{Introduction}
High temperature superconductivity is an important area of research in physics, not only due to its potential applications but also the interesting physics of the strong electron correlation in high-$T_c$ materials. However, the physics of cuprate high-$T_c$ superconductors is still largely unclear \cite{keimer2015quantum}. For example, the pseudogap regime of the phase diagram, out of which superconductivity emerges, is characterized by an anomalous suppression of electron density of states and a variety of competing and co-existing orders \cite{doi:10.1080/00018730903122242}. The precise relationship between these orders and their connection to the pseudogap, and ultimately the mechanism for unconventional superconductivity, remains controversial. Likely, they each play an important, but complex role in the mechanism of superconductivity and are therefore deserving of continued investigation \cite{RevModPhys.75.1201,doi:10.1063/1.4818402}.

Common to the cuprates are charge and spin order in the form of stripes, unidirectional charge- and spin-density waves in the copper oxide plane that break rotational and translational symmetry. While evidence for charge order has been found in all families of cuprates \cite{doi:10.1146/annurev-conmatphys-031115-011401}, there appears to be less universality in spin stripes, with varied behavior in different compounds, such as the periodicity of the modulation and their static or fluctuating nature.  Whether these stripes compete or cooperate with superconductivity still represents an active area of research \cite{RevModPhys.87.457}. A related order, electronic nematicity \cite{kivelson1998electronic}, unlike stripes, breaks rotational symmetry while preserving translational symmetry. The kind of experimental anisotropies suggestive of electronic nematicity have been observed in a number of experiments including transport \cite{Daou2010,Sato2017,Murayama2019,wu2017spontaneous}, neutron scattering \cite{Hinkov597}, scanning tunneling microscope \cite{lawler2010intra} and nuclear magnetic resonance \cite{wu2015incipient}. However, properties, such as the inferred nematic orientation and onset temperature lack any universal character across cuprate families. Electronic nematicity and stripes may be closely related, as nematicity can arise naturally from the melting of stripe order \cite{kivelson1998electronic}. Electron nematicity also may play an important role in the formation or enhancement of superconducting order, as suggested by several theories \cite{Kim_2004,PhysRevLett.114.097001,PhysRevB.91.115111}.

Theoretically, the Hubbard model is considered a canonical starting-point for the study of strongly correlated electrons, and, in particular, has been able to capture signatures of a number of the orders relevant to the cuprates. The case for stripes was made by early mean-field calculations, which predicted their formation \cite{PhysRevB.40.7391,MACHIDA1989192,doi:10.1143/JPSJ.59.1047}, and subsequently more sophisticated methods have further corroborated the presence of stripes in both the ground state and at finite temperatures within the Hubbard model~\cite{PhysRevLett.80.1272,PhysRevB.71.075108,PhysRevLett.104.116402,PhysRevB.92.205112,Zheng1155,Huang1161,huang2018}. While a number of methods have been developed to study strongly correlated models, including the Hubbard model, a particularly powerful, finite-temperature, numerically exact technique is determinant quantum Monte Carlo (DQMC)~\cite{PhysRevB.40.506}, although it has been restricted to relatively high temperatures due to the fermion sign problem~\cite{PhysRevB.92.045110,PhysRevB.41.9301,PhysRevLett.94.170201}. Recent DQMC simulations have shown the presence of fluctuating spin stripes, but with no discernible signatures of fluctuating charge order, possibly due to the high simulation temperatures~\cite{Huang1161,huang2018}. This leads us to ask whether signatures of fluctuating charge order may be present in a different form in the Hubbard model at the temperatures accessible by DQMC and whether these may be corroborated by experiment. 

Experimentally, the nematic susceptibility is proportional to the real part of the Raman response at zero frequency (static Raman susceptibility) \cite{auvray2019nematic, PhysRevLett.111.267001}.  This will enable us to compare the results from simulations directly against Raman measurements on cuprates. Previous Raman spectra that were taken on La\textsubscript{2-x}Sr\textsubscript{x}CuO\textsubscript{4} (LSCO) have shown prominent low energy peaks at low temperatures due to charge stripe excitations \cite{muschler2010electron}. In underdoped cuprates this contribution dominates the spectra, since particle-hole excitations in $B_\mathrm{1g}$ symmetry are largely gapped out below 1000\,$\mathrm{cm}^{-1}$ in the pseudogap~\cite{PhysRevLett.95.117002}. A key question will be whether similar signatures can be found in simulations of the Hubbard model, even at high temperatures, and how such features can be interpreted in terms of electronic nematic fluctuations and ``melted'' stripes.

Here, using DQMC we calculate the nematic susceptibility of the single-band Hubbard model on a square lattice, which is proportional to the static Raman response, and draw a comparison to Raman scattering experiments in LSCO. The nematic susceptibility in $B_{1g}$ symmetry of the Hubbard model, with no explicit symmetry breaking terms, shows an unexpected non-monotonic dependence on doping, not tied to simple band structure effects, such as the van Hove singularity, but rather this may arise from strong electron correlations. The Raman $B_{1g}$ susceptibility, extracted from the Raman spectra of LSCO, shows a similar, but more dramatic doping dependence, albeit at much lower temperatures, which previously has been attributed to charge excitations in the presence of stripes \cite{PhysRevB.66.060502,PhysRevLett.95.117002}. 

\section{Methods}
The Hubbard model is one of the simplest models describing strongly correlated electron physics and has been taken as a starting point for understanding the low-energy physics of the cuprates~\cite{PhysRevB.37.3759}. The Hamiltonian is given by
\begin{equation}
    H = -\sum_{ij\sigma} t_{ij} c_{i\sigma}^\dagger c_{j\sigma} + U \sum_{i}\hat{n}_{i\uparrow}\hat{n}_{i\downarrow} - \mu\sum_{i\sigma}\hat{n}_{i\sigma}, \label{eq:hub_model}
\end{equation}
where $t_{ij}$ are electron hopping matrix elements which parametrize the kinetic energy (here, we assume only non-zero nearest- and next-nearest-neighbor hopping matrix elements); $U$ is the Hubbard repulsion, giving rise to the strongly correlated nature of the model; $\mu$ is the chemical potential controlling the number of electrons; $c^{\dagger}_{i\sigma}$ ($c_{i\sigma}$) are electron creation (annihilation) operators for electrons at site $i$ with spin $\sigma$; and $\hat{n}_{i\sigma} = c^\dagger_{i\sigma}c_{i\sigma}$ is the number operator. As in previous studies of stripes~\cite{Huang1161}, we choose $U = 6t$ to ensure that we capture the effects of strong correlations while simultaneously mitigating the effects of the fermion sign problem, which allows us to access lower temperatures in our DQMC simulations. The results that we present below were obtained from 64-site square clusters.

In our numerical simulations, the Raman response in imaginary time $s(\tau)$ is given by the correlation function of the effective scattering operator $\rho_{\alpha}$~\cite{RevModPhys.79.175}
\begin{equation}
    s_{\alpha}(\tau) = \langle T_\tau \rho_{\alpha}(\tau)
    \rho_{\alpha}^\dagger(0) \rangle, \label{eq:corr_fn}
\end{equation}
where $\alpha$ is one of the two common symmetry projected scattering channels $B_{1g}$ or $B_{2g}$ and the scattering operators are properly projected charge density operators.  These projected symmetry channels highlight different portions of the Brillouin zone and depend on the polarization of the incident and scattered light in a Raman experiment, such that the scattering operators in momentum space take the form
\begin{equation}
    \rho_{B_{1g}} = \frac{1}{2}\sum_{\boldsymbol{k}\sigma}(\cos k_x - \cos k_y) c^\dagger_\sigma(\boldsymbol{k})c_\sigma(\boldsymbol{k}),
\end{equation}
\begin{equation}
    \rho_{B_{2g}} = \sum_{\boldsymbol{k}\sigma} \sin k_x \sin k_y c^\dagger_{\sigma} (\boldsymbol{k}) c_{\sigma}(\boldsymbol{k}).
\end{equation}
These operators, and the corresponding response function, can capture broken $C_4$ rotational symmetry, or nematicity, as one can see when the operators are written in real space
\begin{equation}
    \rho_{B_{1g}} = \sum_{i\sigma} \big(c^\dagger_{i,\sigma}c_{i+\hat x,\sigma} - c^\dagger_{i,\sigma}c_{i+\hat y,\sigma} + \text{h.c.}\big), \label{eq:hub_b1g}
\end{equation}
\begin{equation}
    \rho_{B_{2g}} = \sum_{i\sigma} (c^\dagger_{i,\sigma}c_{i + \hat x + \hat y,\sigma} - c^\dagger_{i,\sigma}c_{i - \hat x + \hat y,\sigma} + \text{h.c.}). \label{eq:hub_b2g}
\end{equation}
Real-frequency Raman spectra can be obtained by analytic continuation, equivalent to inverting the following equation
\begin{equation}
    s_{\alpha}(\tau) = \langle T_\tau \rho_{\alpha}(\tau)\rho_{\alpha}^\dagger \rangle = \int_{-\infty}^\infty \frac{d\omega}{\pi} \frac{e^{-\tau\omega}}{1-e^{-\beta\omega}} \chi_{\alpha}''(\omega), \label{eq:ac}
\end{equation}
where $\chi_{\alpha}''(\omega)$ is the imaginary part of the Raman response, directly comparable to Raman measurements. Inverting Eq. (\ref{eq:ac}) is ill-defined because of the behavior of the kernel at large frequencies; however, after integrating over imaginary time
\begin{equation}
    \int_0^\beta d\tau\, s_{\alpha}(\tau) = 2\int_{0}^\infty \frac{d\omega}{\pi} \frac{\chi_{\alpha}''(\omega)}{\omega} \equiv \chi_{\alpha}'(\omega=0). \label{eq:int_ram_sus}
\end{equation}
The left-hand side of this expression can be determined entirely from the results of DQMC simulations. The right-hand side is just a Kramers-Kronig relation,  proportional to the real part of the Raman response function at zero frequency $\chi_{\alpha}'(\omega=0)$, which can be readily obtained from experimental measurements, allowing for a direct comparison.

\section{Results}
\begin{figure}
    \includegraphics[scale=0.5]{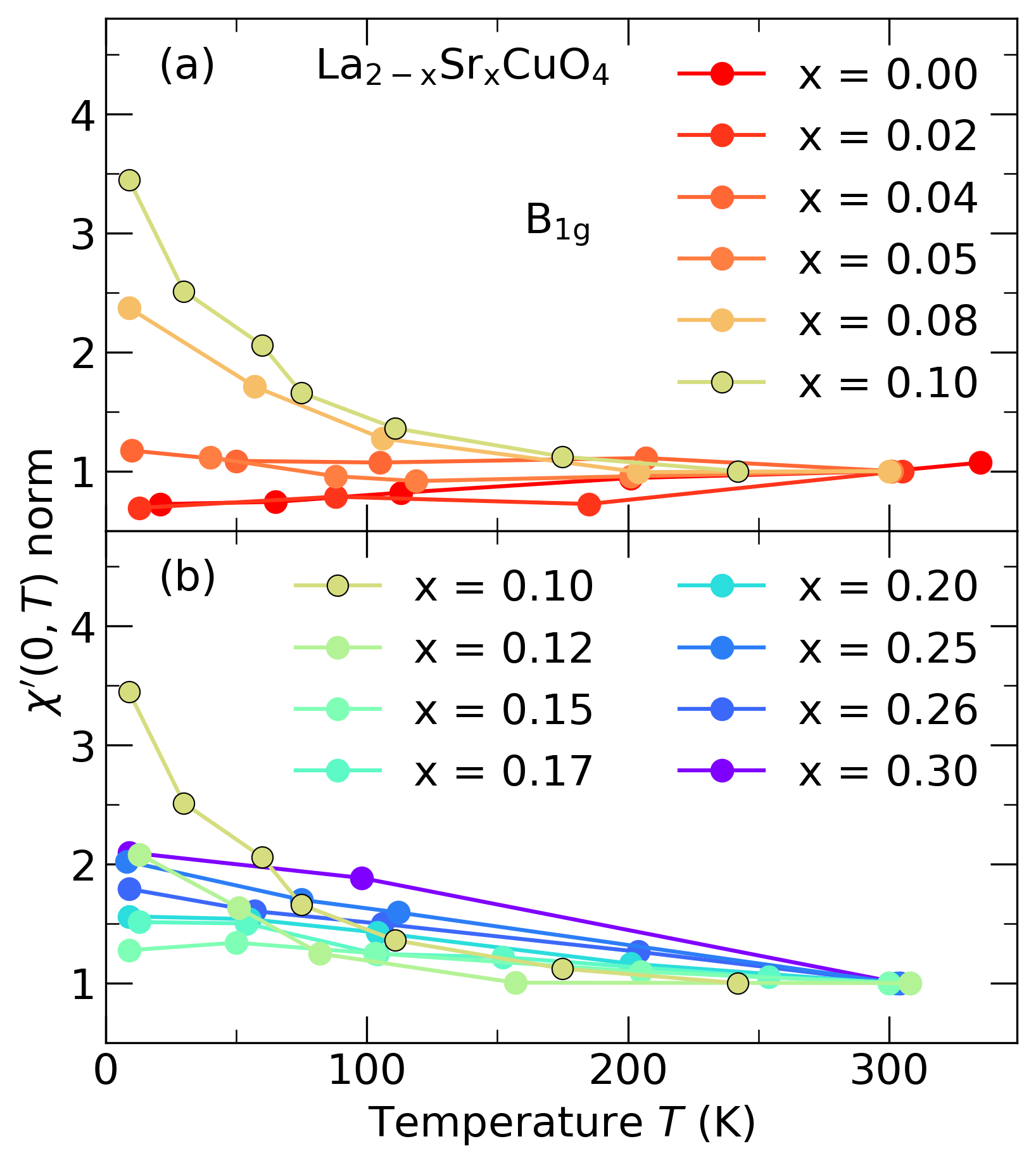}%
    \caption{Real part $\chi^\prime_{B1g}(T)$ of the static Raman susceptibility in $B_{1g}$ symmetry as a function of temperature for various doping levels $x$. The doping ranges below and above $x=0.1$ are shown separately for clarity in (a) and (b), respectively. A clear Curie-like increase of $\chi^\prime_{B1g}(T)$ at low temperatures can only be resolved for $x=0.08,0.1$ and $0.12$. }
    \label{fig:chiB1gT}
\end{figure}

Published work from earlier Raman scattering studies on LSCO showed that a low energy peak in $B_{1g}$ symmetry emerged with decreasing temperature, most prominently near $10\%$ hole doping~\cite{muschler2010electron}, attributed to charge stripe excitations~\cite{PhysRevLett.95.117004}. The real part of the Raman response at zero frequency (static Raman susceptibility) is dominated by these low energy features as the integrand falls off as $1/\omega$.
Fig.~\ref{fig:chiB1gT} shows data for $\chi^\prime_{B1g} (T)$ determined from previous experimental spectra~\cite{muschler2010electron}. $\chi^\prime_{B1g} (T)$ exhibits a Curie-Weiss-like temperature dependence for the doping concentrations $x=0.08,\, 0.1$ and $0.12$. At higher doping there is an increase of  $\chi^\prime_{B1g} (T)$ as temperature decreases, but the shape is convex rather than concave. In the overdoped regime, the particle-hole continuum represents a dominant contribution to the intensity and any possible contributions from stripe fluctuations cannot be resolved, and presumably are not present~\cite{muschler2010electron}. 

\begin{figure}
    \includegraphics[scale=0.5]{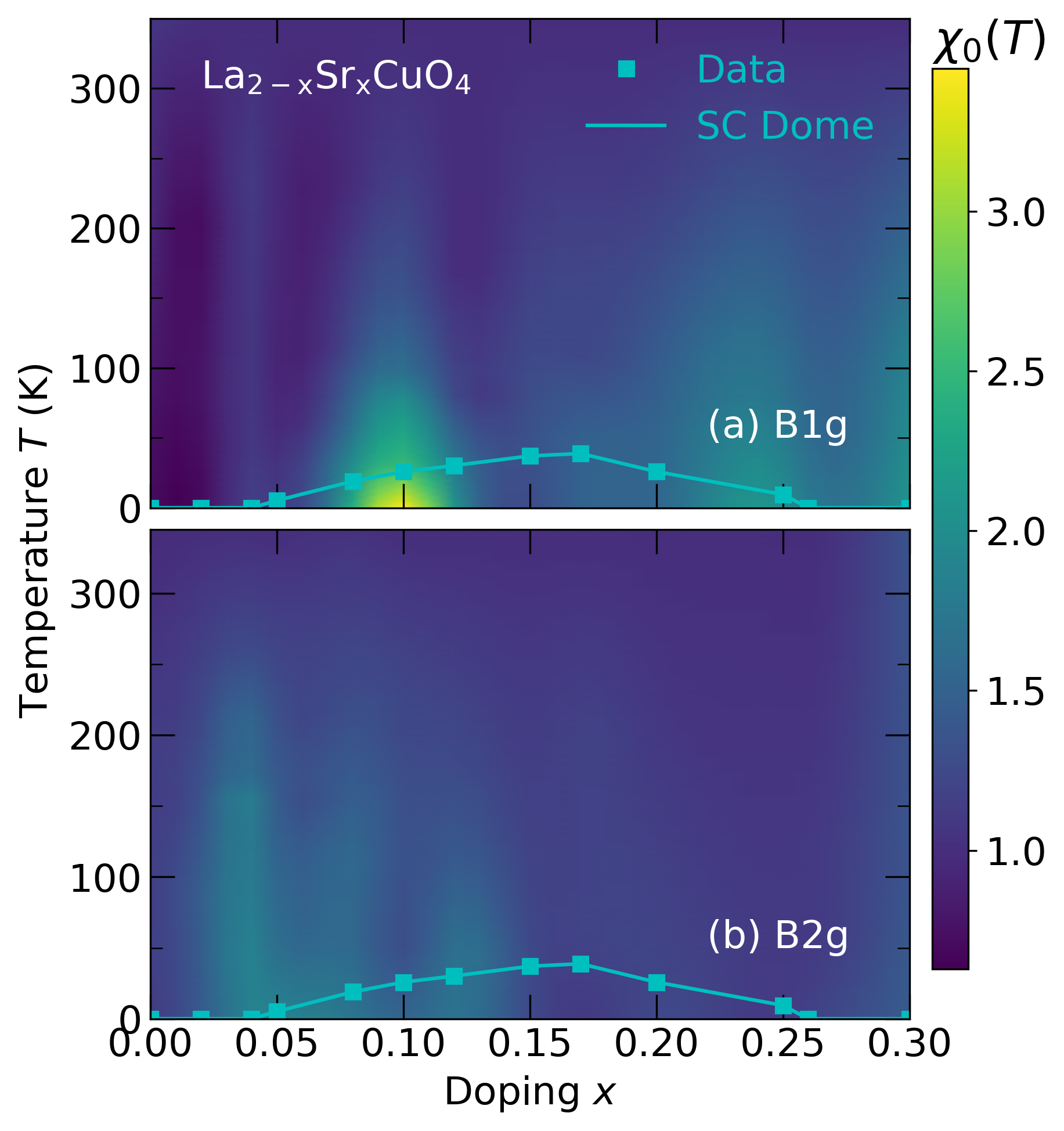}%
    \caption{Continuous color plots of the susceptibility $\chi^\prime(x,T)$. (a) In $B_{1g}$ symmetry $\chi^\prime(T)$ increases in the region around $x=0.1$. (b)  In comparison, the $B_{2g}$ susceptibility is weakly temperature dependent at the onset of superconductivity and remains flat otherwise.}
    \label{fig:lsco}
\end{figure}

To highlight the special nature of $x=0.1$, we interpolate $\chi^\prime_{B1g}(x,T)$ from the experimental data and plot an $x$-$T$ ``phase" diagram in Fig. \ref{fig:lsco}. $\chi^\prime_{B1g} (x,T)$ is enhanced prominently in the vicinity of $x=0.1$ at low temperatures and falls off at higher and lower doping. A continuous increase towards higher doping originates from the buildup of the particle-hole continuum. In contrast to the substantial temperature dependence of the susceptibility in $B_{1g}$ around $x=0.1$, there is only a small increase in the otherwise flat response of $\chi^\prime_{B2g} (T)$ at the onset of the superconducting dome [\textit{cf.} Fig.~\ref{fig:lsco} (b) and in the Appendix (Additional Figures) Fig. \ref{fig:chiB2gT}, which shows the real part of the Raman susceptibility in the $B_{2g}$ channel at the corresponding doping values].

\begin{figure}
    \includegraphics[scale=0.5]{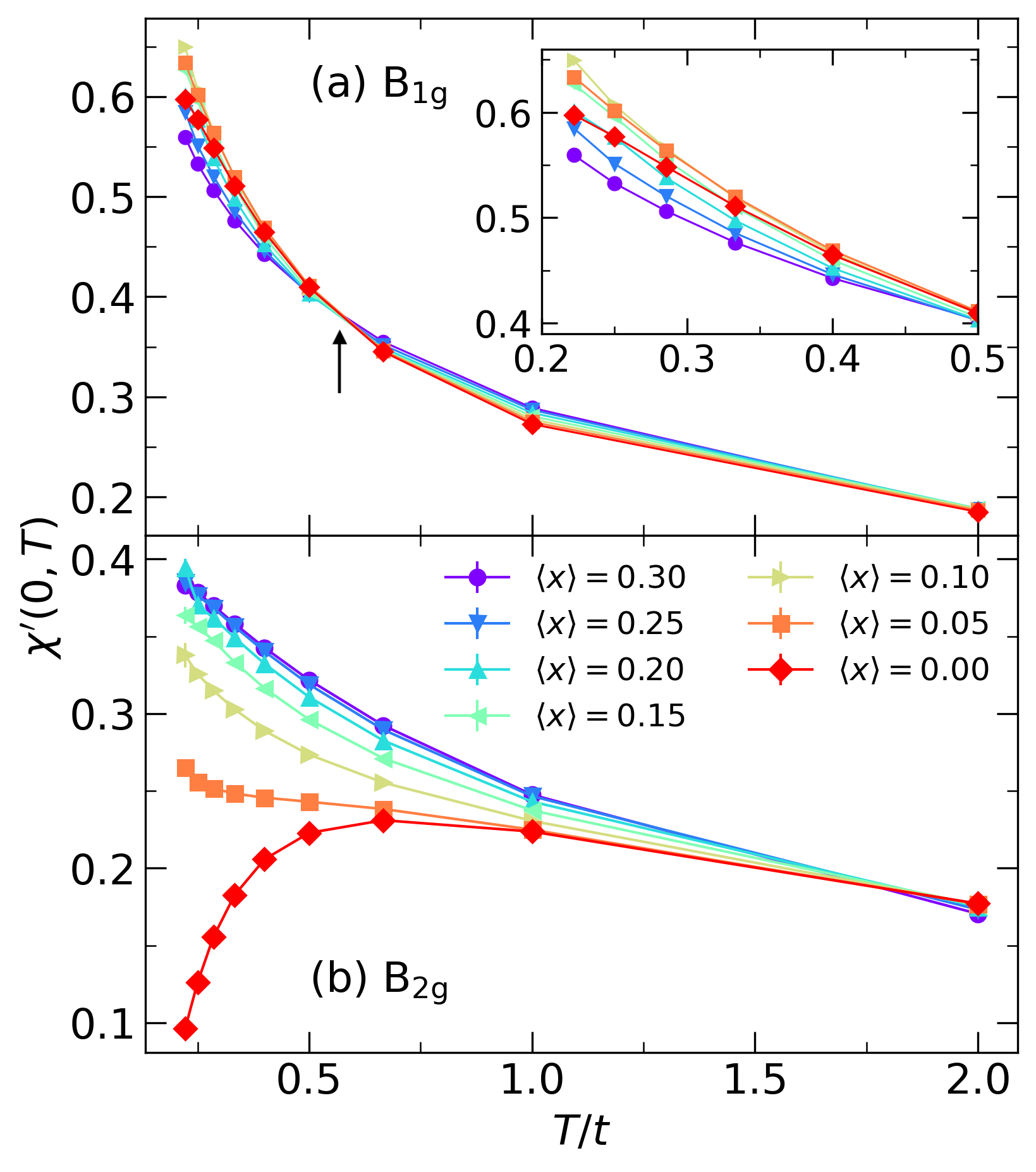}%
    \caption{Nematic/Raman susceptibility calculated using DQMC as a function of temperatures for various doping levels for (a) $B_{1g}$ symmetry and (b) $B_{2g}$ symmetry. Both susceptibility increases with decreasing temperature. For $B_{1g}$, the rate of increase for different doping becomes different as temperature is lowered through the point indicated by the arrow, which is near $T=2t/3=J$ (the data point to the right of the arrow), the magnetic exchange. The $B_{2g}$ susceptibility does not show similar behavior, and is gapped at half-filling.}
    \label{fig:qmc1}
\end{figure}

\begin{figure}
    \includegraphics[scale=0.5]{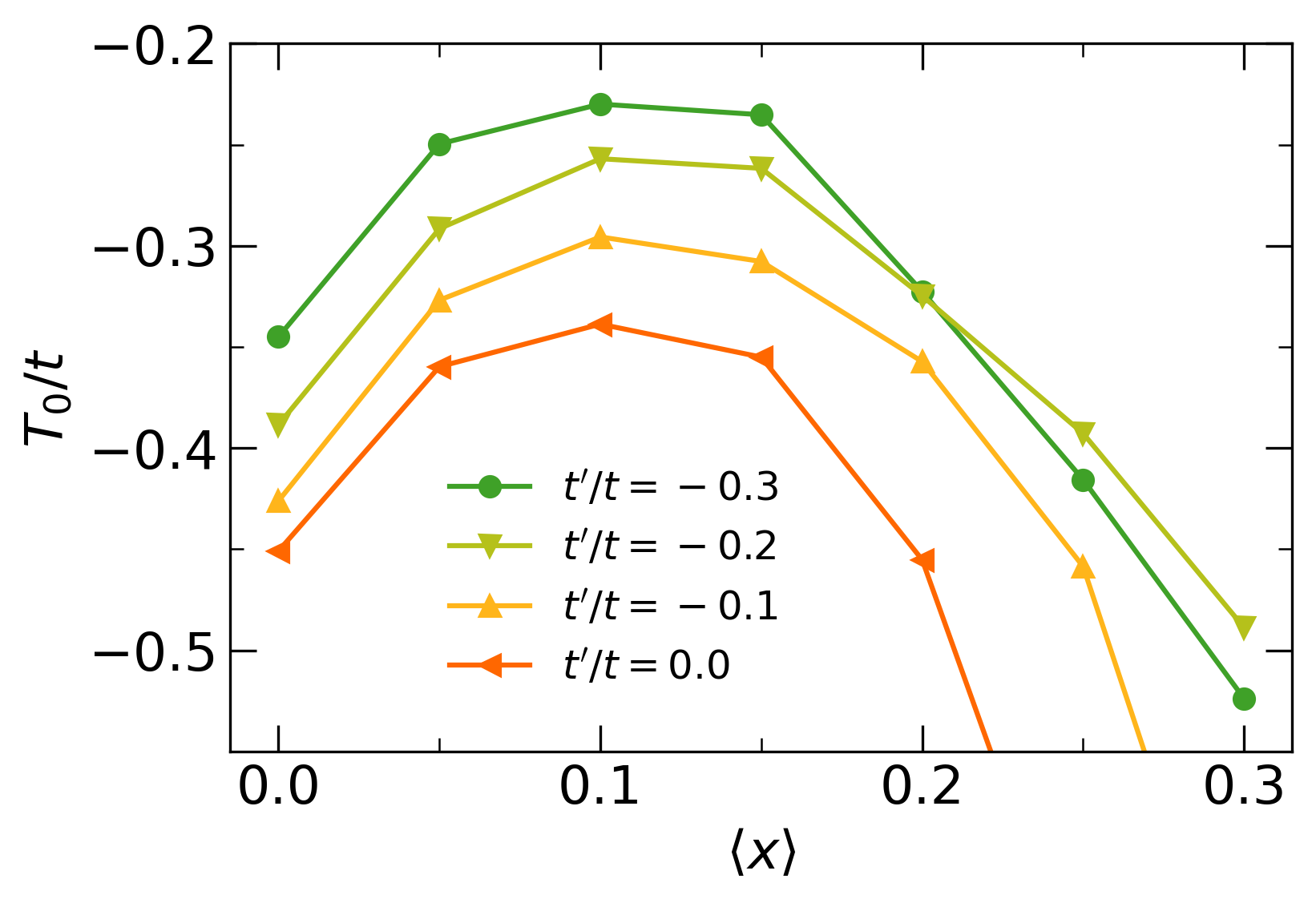}%
    \caption{Curie-Weiss temperature $T_0$ obtained from Curie-Weiss fit of the $B_{1g}$ susceptibility as a function of doping level for different values of $t'$, all showing a peak at $\langle x \rangle=0.1$. The peak moves slightly to higher doping level with increasing $-t'/t$, which may be due to the van-Hove point moving to higher doping level with increasing $-t'/t$.}
    \label{fig:Tc_b1g_tps}
\end{figure}

To address these findings, we calculate the nematic susceptibility of the Hubbard model [Eq. (\ref{eq:int_ram_sus})] using DQMC. The susceptibility as a function of temperature for various doping levels is displayed in Fig. \ref{fig:qmc1}. We choose a next-nearest-neighbor hopping of $t'/t=-0.25$ to mimic a similar Fermi surface to that of the cuprates. However, and as we show later, the behavior that we observe is not tied to any special features of the band structure, such as the position of the van Hove singularity that would be tied to this choice of $t'$ [Fig. \ref{fig:Tc_b1g_tps}].  As one can see from Fig. \ref{fig:qmc1}, the general trend for the $B_{1g}$ susceptibility is to increase with decreasing temperature, indicating an enhancement of the nematic correlations.  This would be expected on general grounds as reduced thermal fluctuations at lower temperatures result in sharper charge excitation peaks and an enhanced susceptibility. Unlike the susceptibility extracted from experiments at half-filling, which decreases with temperature due to the presence of a gap at low frequency, the susceptibility derived from DQMC simulations does not decrease. This can be attributed to contributions from relatively broad two-magnon excitations, which extend to low energy at DQMC temperatures, expected to manifest in the $B_{1g}$ symmetry channel. The two-magnon peak in experiments, while present, is comparatively sharp and at too high a relative energy to contribute significantly to the susceptibility $\chi^\prime_{B1g} (T)$.

The rate of increase for different doping levels becomes different as the temperature is lowered through the value indicated by the arrow, which is near $T = 2t/3 =J$, the magnetic exchange for the Hubbard model with the chosen parameters. Although the increase of the susceptibility going from the highest temperature to the lowest is about three times for all dopings, it is clear that at low temperatures the rate of increase has a non-monotonic dependence on doping, as shown in the zoomed-in view in the inset. In particular, at $\langle x \rangle = 0.1$, the susceptibility increases the fastest, qualitatively similar to the behavior in experiments, where the Raman susceptibility is highly enhanced at this doping. We note that at half-filling, the curve of the susceptibility turns from convex to concave at low temperatures. At finite temperatures, interband transitions at frequencies smaller than the onset of the Mott gap are present. As temperature is lowered, such transitions are suppressed, giving rise to the smaller rate of increase for the susceptibility.

The $B_{2g}$ nematic susceptibility is plotted in Fig. \ref{fig:qmc1} (b). At half-filling, unlike $B_{1g}$, the susceptibility decreases with temperature. Since the two-magnon excitation is mostly absent in the $B_{2g}$ symmetry, the susceptibility is determined predominantly by the Mott gap of the half-filled Hubbard model. The $B_{2g}$ Raman spectral weight at small frequencies due to finite temperatures are suppressed as temperature is lowered, resulting in decreasing susceptibility. For all other dopings, the susceptibility rises with decreasing temperature, similar to $B_{1g}$. However, the rate of increase does not show a non-monotonic dependence on the doping level as seen in $B_{1g}$. Instead, the susceptibility rises with increasing hole doping monotonically, due to the increasing spectral weight of the quasiparticles in the Brillouin zone diagonal.

At low temperatures, the $B_{1g}$ susceptibility is approximately inversely proportional to temperature, suggesting that it can be fit to a Curie-Weiss form. The Curie-Weiss temperature, $T_0$, as a function of doping for different values of the next-nearest neighbor hopping $t'$ is shown in Fig. \ref{fig:Tc_b1g_tps}. While $T_0$ is negative in all cases, suggesting that the system does not possess a transition into an ordered nematic state, at least as far as one can tell at these temperatures, it does show a non-monotonic dependence on doping. While broad, the maximum of $T_0$ occurs around $\langle x \rangle = 0.1$ for all values of $t'$.  This behavior generally agrees with the experimental data which shows a sharper, more prominent peak at a similar doping.

The results presented here were all obtained from 64-site square clusters. As shown in the Appendix (Additional figures) Fig.~\ref{fig:N10_8_Tcs}, we also have checked that the notable behavior is present in 100-site square clusters.  While finite-size effects clearly are present in the data, the peak in $\chi^\prime_{B1g} (T)$ remains, both qualitatively and quantitatively.

\section{Discussion}
Although DQMC calculations are performed at high temperatures, many of the properties of the Hubbard model obtained with this method are still consistent with those found in experiments done at low temperatures on cuprate materials. Here, our results show that the doping dependence of the Raman susceptibility in LSCO is captured by the Hubbard model. The enhancement of the susceptibility at $\langle x \rangle = 0.1$ in LSCO is much more prominent than that seen in the calculations, because of the low temperature of the experiments. We expect this similar enhancement in calculations to be more pronounced as temperature is lowered.

The maximum of the $B_{1g}$ nematic susceptibility may be closely related to fluctuating stripes. Electron nematicity may arise from melting of stripes due to quantum or thermal fluctuations, where the translational symmetry is restored but the rotational symmetry is still broken \cite{kivelson1998electronic}. It is possible that at the temperatures accessible by DQMC, charge stripe has been melted, leaving an electron fluid with a nematic correlation that is strongest near the doping where stripe correlation is otherwise strong. Furthermore, the enhancement of the $B_{1g}$ Raman susceptibility of LSCO due to prominent low energy peaks in the Raman spectra may be interpreted in terms of signals for charge excitations within stripes \cite{PhysRevLett.95.117002,PhysRevB.66.060502,PhysRevLett.95.117004}. This further corroborates the connection between electron nematicity and stripes in cuprates and the Hubbard model.

The Raman response which relates to the nematic susceptibility through the Kramers-Kronig relation may be obtained from the imaginary-time data using analytic continuation. Here we use the maximum entropy method (MEM) to perform numerical analytic continuation \cite{JARRELL1996133}. Although the resulting spectra are generally broad due to the high temperatures of the simulation, some qualitative comparisons may be made between the $B_{1g}$ and $B_{2g}$ responses, and between theory and experiments.

\begin{figure}
    \includegraphics[scale=0.6]{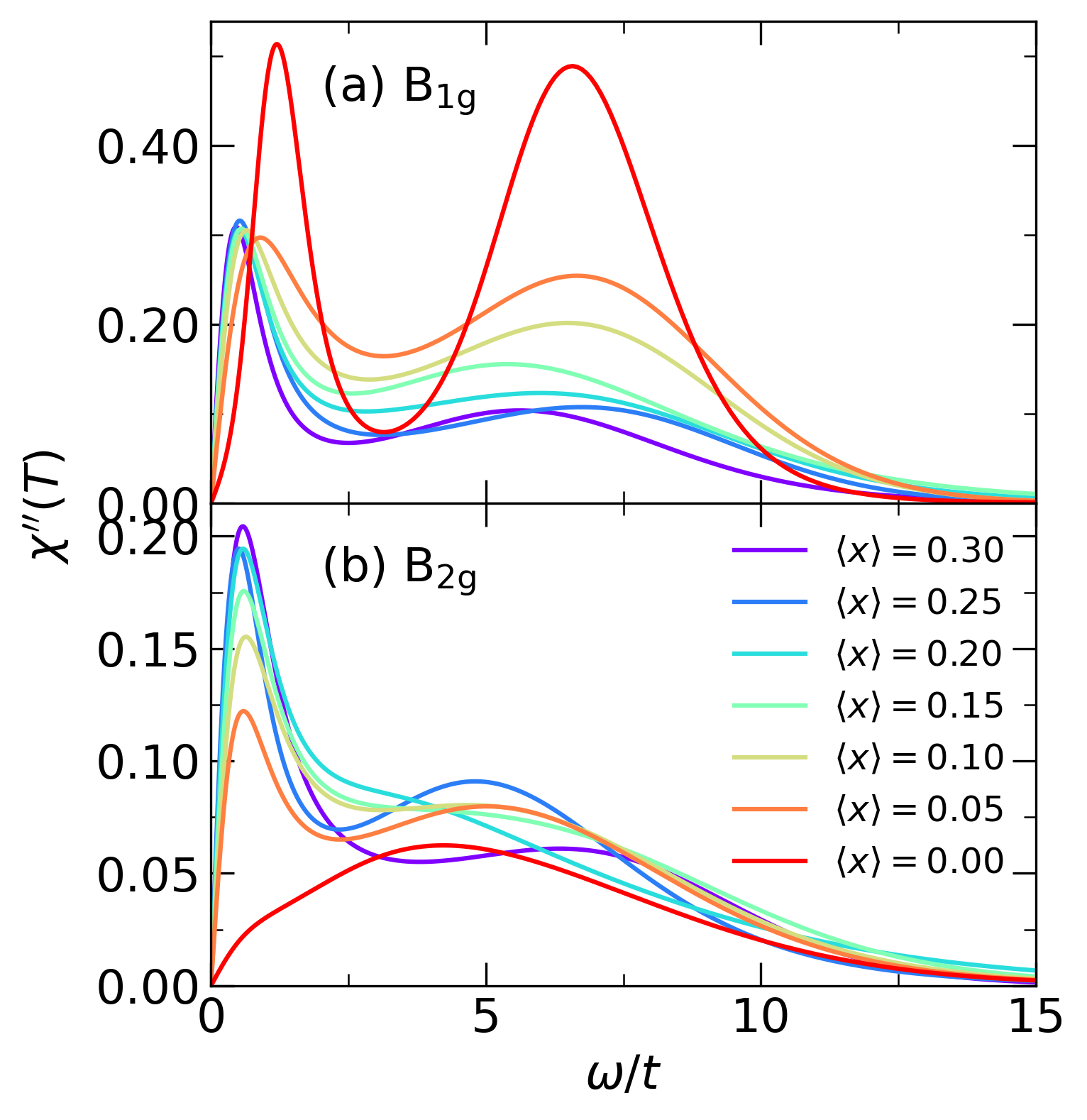}%
    \caption{Raman spectra obtained via maximum entropy analytic continuation for (a) $B_{1g}$ and (b) $B_{2g}$ symmetries for different doping levels, at $\beta = 4.5/t$ and $t'=-0.25t$. The broad peak at around $\omega/t=6$ arises from interband transitions across the Mott gap.}
    \label{fig:ram_spectra}
\end{figure}

Fig. \ref{fig:ram_spectra} shows the spectra for different doping levels at $T/t=0.22$. At half-filling, the $B_{1g}$ spectrum (a) shows two peaks, one at a frequency slightly higher than $\omega/t=6=U/t$, and one at around $\omega/t=1.2$. By contrast, the $B_{2g}$ spectrum (b) shows only one broad peak. The low energy feature in the $B_{1g}$ spectrum can be attributed to the two-magnon scattering, which is not captured by the $B_{2g}$ form factor. With hole doping, the high energy peak in the $B_{1g}$ spectrum broadens, and the peak height decreases with doping due to spectral weight being transferred to the quasiparticle band that appeared with hole doping \cite{PhysRevB.84.235114}. For the low energy part, the two-magnon peak is quickly reduced when doped away from half-filling, and a quasiparticle peak at lower energy appears. With further hole doping, the quasiparticle peak sharpens while maintaining the peak height. For $B_{2g}$, the high energy part shows varying peak heights and positions for different doping levels, but remains generally broad. At low energy the $B_{2g}$ spectrum develops a clear quasiparticle peak which becomes more prominent with hole doping.

The $B_{1g}$ and $B_{2g}$ form factors highlight the anti-nodal and nodal regions of the Brillouin zone, respectively. With doping, the quasiparticles in the anti-nodal region are less coherent than those in the nodal region, due to the much stronger quasiparticles scattering, which have been also observed in ARPES experiments \cite{PhysRevB.78.205103}. Since the Kramers-Kronig transform highlights the low energy part [Eq. (\ref{eq:int_ram_sus})], the steady increase of the coherent quasiparticle peak in $B_{2g}$ with doping gives rise to the monotonically increasing $B_{2g}$ susceptibility. On the other hand, the quasiparticle peak in $B_{1g}$ does not increase with doping. The apparent shift of the low energy part to lower energy, when doped from half-filling to about $\langle x \rangle = 0.1$ gives rise to the initial increase of the $B_{1g}$ susceptibility. Upon further doping, the relatively unchanged quasiparticle peak and the decreasing high energy part together lead to the decrease in the susceptibility, and hence a maximum at $10\%$ hole doping.

Detailed studies of the Raman spectra of LSCO have been reported previously \cite{muschler2010electron}. While the experimental spectra and the theoretical ones presented in Fig. \ref{fig:ram_spectra} cannot be compared quantitatively, certain features of the spectra are consistent. In particular, in both LSCO and the Hubbard model, the dominant peak in the $B_{1g}$ spectra are from the two-magnon excitation (compare Fig. 8 in \cite{muschler2010electron} and Fig. \ref{fig:ram_spectra}). A previous study on the Raman response in the Hubbard model using DQMC did not produce the two-magnon peak, because the form of the response function used did not include vertex corrections and hence only captured particle-hole excitations projected by the Raman form factors \cite{PhysRevB.84.235114}. Here, the full Raman response function goes beyond particle-hole excitations and includes contributions from higher order processes that give rise to the two-magnon scattering. This demonstrates the ability of the Hubbard model, which describes electronic degrees of freedom, to capture an important feature in the cuprates due to spin excitations. In addition, the experimental $B_{1g}$ response develops a peak from electron-hole excitations at high doping, which has similar trends in the theoretical spectra.

Muschler et al. discussed several ways of analyzing the Raman spectra, such as calculating the Raman resistivity, $\Gamma_0(T)$, and integrated spectral weight \cite{muschler2010electron}. In particular, in connection to the low energy charge stripe excitations, the Raman resistivity, or the inverse of the initial slope of the spectrum, provides information about the dynamics of the quasiparticles in different symmetry channels. For cuprates with maximum $T_c$ around 100K such as Bi\textsubscript{2}Sr\textsubscript{2}CaCu\textsubscript{2}O\textsubscript{8+$\delta$}, YBa\textsubscript{2}Cu\textsubscript{3}O\textsubscript{y} and Tl\textsubscript{2}Ba\textsubscript{2}CuO\textsubscript{6+$\delta$}, a dichotomy between the $B_{1g}$ and $B_{2g}$ relaxation rates is found at about 20\% doping where $\Gamma_{0,B2g}(T)$ shows metallic behavior while $\Gamma_{0,B1g}(T)$ is $T$-independent. The behavior of LSCO is different in that $\Gamma_{0,B1g}(T)$ increases with $T$ for a wide doping range down to at least 5\%. Muschler et al. indicated that the low energy peaks in the $B_{1g}$ response giving rise to the metallic behavior of $\Gamma_{0,B1g}(T)$ are due to charge excitations in the presence of stripes, which was supported by its consistency with other experimental findings as well as decriptions of microscopic models.

Here we analyzed the Raman spectra by taking the Kramers-Kronig transform, which highlights the low energy part of the spectra dominated by the charge stripe excitations. The qualitatively similar behavior of the obtained Raman (nematic) susceptibility between experiment and theory and the close relation between fluctuating charge stripes and nematicity, provides one more piece of evidence supporting the charge stripe excitation interpretation of the low energy structures in the $B_{1g}$ response of LSCO.

\section{Conclusions}
Using DQMC calculations, we have shown that the doping dependence of the $B_{1g}$ nematic susceptibility of the Hubbard model qualitatively agrees with the Raman $B_{1g}$ susceptibility in LSCO. Specifically, both have a maximum at 10\% hole doping. The peak in the susceptibility from experiments are much more prominent than that in the calculations, likely due to the high temperatures of the calculations. We expect the nematic fluctuations to continue to grow as temperature is lowered beyond the region accessible to DQMC simulaitons, as can be seen from the Curie-Weiss temperatures in Fig. \ref{fig:Tc_b1g_tps}.

The agreement between the calculations and the experiments is consistent with the picture of electron liquid crystals. In this framework, melting of charge stripes due to quantum or thermal fluctuations may give rise to electron nematicity. Since DQMC calculations of the Hubbard model is performed at very high temperatures, it is possible that charge stripes are melted and result in a state with strong nematic correlation. The doping where the nematic correlation of the Hubbard model is maximum coincides with the doping where charge stripe excitations in LSCO gives the most prominent Raman signal, supporting the scenario of the electron liquid crystals.

While the theoretical spectra from analytic continuation cannot be quantitatively compared to the experimental spectra, the prominent two-magnon peak in the $B_{1g}$ spectra at half-filling and its quick reduction upon doping, as well as the emerging quasiparticle peak, are observed in both experiment and theory. Previous work by Muschler et al. discussed several analyses, such as Raman resistivity and integrated spectral weight, in connection to the low energy structures due to charge stripe excitations. In this work we analyzed the spectra by calculating its real part at zero frequency through the Kramers-Kronig relation, and the comparison with thoretical calculations further supports the attribution of charge stripes to the emerging low energy structures in the spectra.

Our result demonstrates the capability of the Hubbard model to capture tendencies of electron nematicity that aligns with experiments in cuprates. The agreement with experiments also supports the notion of electron liquid crystal states where nematicity results from melting of stripe order.

\section{Acknowledgments}
T.L. acknowledges helpful discussions with Y.-F. Jiang. This work was supported by the US Department of Energy, Office of Basic Energy Sciences, Materials Sciences and Engineering Division, under Contract No. DE-AC02-76SF00515 (T.L., B.M., E.W.H and T.P.D for theoretical and computational work). E.W.H. was supported by the Gordon and Betty Moore Foundation EPiQS Initiative through the grants GBMF 4305 and GBMF 8691. The experimental work in Garching (D.J. and R.H.) was supported by the Bavaria-California Technology Center (BaCaTeC) under Grant No. 21[2016-2] and the German Research Foundation (DFG), Project IDs 107745057 - TRR80 and HA2071/12-1. D.J. acknowledges support from the Friedrich-Ebert-Stiftung. Computational work was performed on the Sherlock cluster at Stanford University and on resources of the National Energy Research Scientific Computing Center, supported by the U.S. Department of Energy under contract DE-AC02-05CH11231.

\bibliography{ramanbib}

\onecolumngrid
\newpage
\appendix*
\section{Additional figures}
\begin{figure}[h]
    \centering
    \begin{minipage}{0.48\textwidth}
        \includegraphics[scale=0.5]{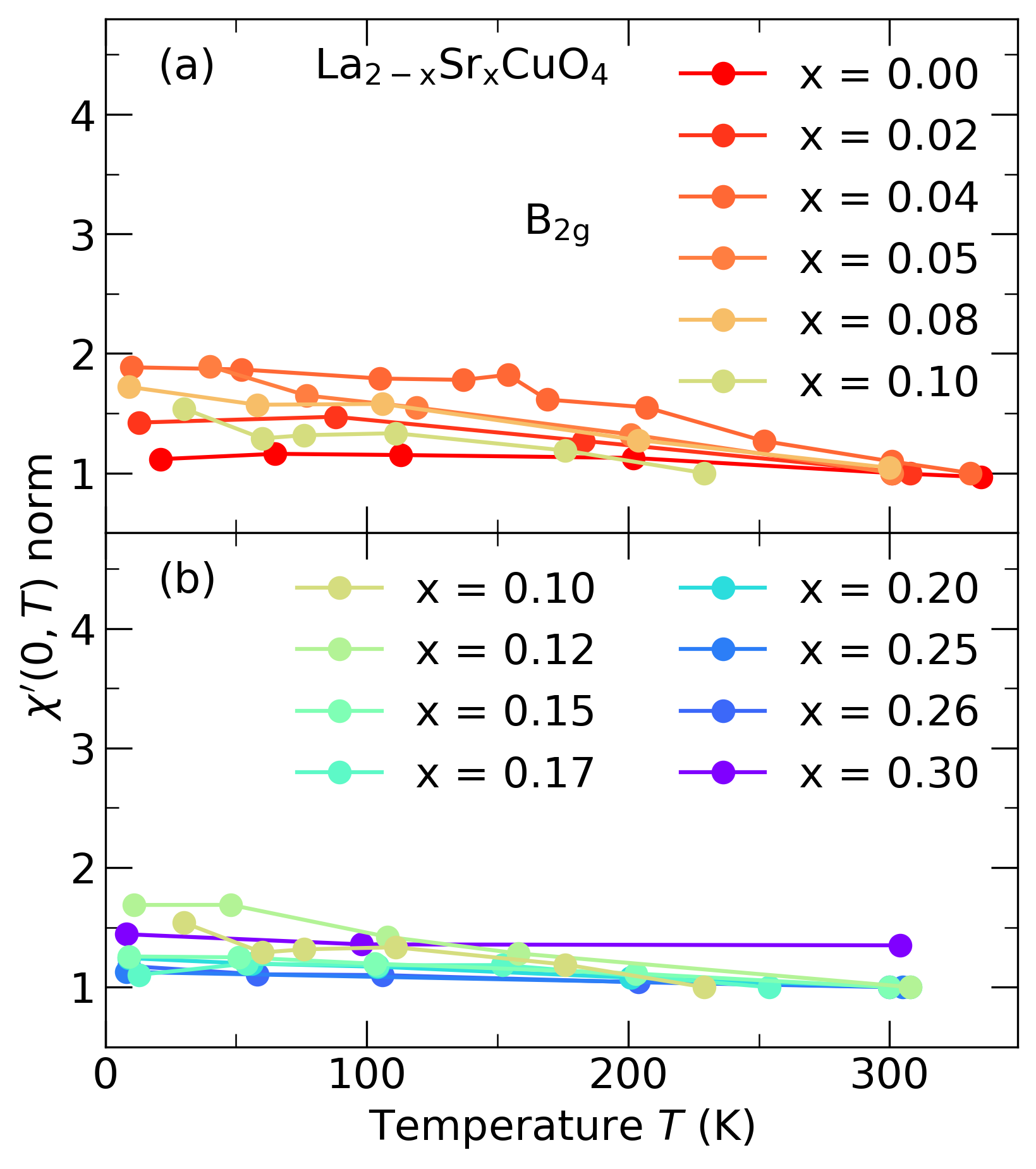}
        \caption{Real part $\chi^\prime_{B2g}(T)$ of the static Raman susceptibility in $B_{2g}$ symmetry as a function of temperature for various doping levels $x$. The doping ranges below and above $x=0.1$ are shown separately for clarity in (a) and (b), respectively.}%
        \label{fig:chiB2gT}
    \end{minipage}\hfill
    \begin{minipage}{0.48\textwidth}
        \includegraphics[scale=0.5]{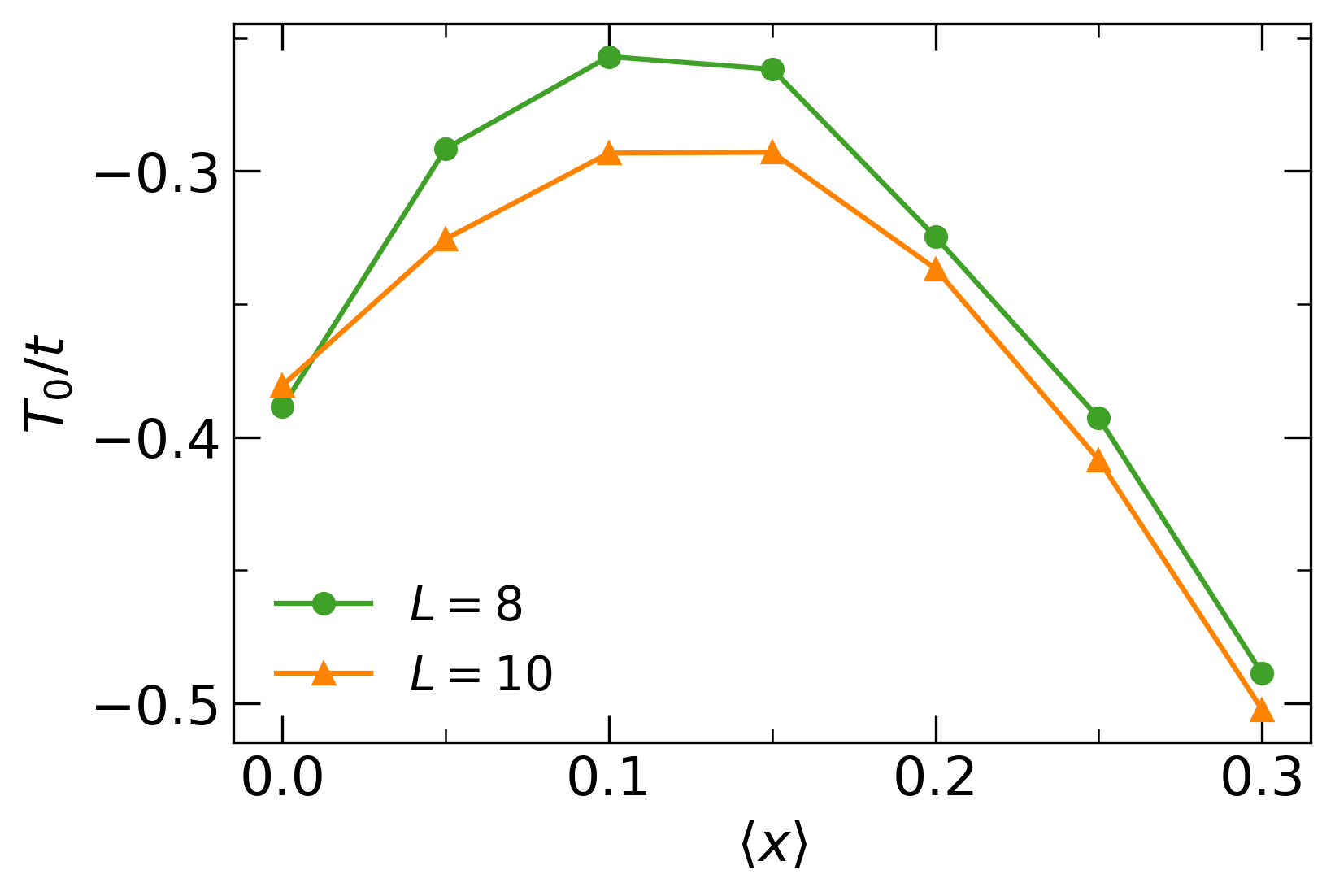}
        \caption{Curie-Weiss temperature $T_0$ for the $B_{1g}$ susceptibility as a function of doping, for $L=8$ and $L=10$ square clusters, both showing a peak around $\langle x \rangle = 0.1$. The next-nearest-neighbor hopping is $t'=-0.2t$.}%
        \label{fig:N10_8_Tcs}
    \end{minipage}
\end{figure}

\end{document}